\documentclass{JHEP}
\usepackage{epsf}

\newcommand{\onefigure}[2]{\begin{figure}[htbp]
         \caption{\small #2\label{#1}(#1)}
         \end{figure}}

\renewcommand{\onefigure}[2]{\begin{figure}[htbp]
         \begin{center}\leavevmode\epsfbox{#1.eps}\end{center}
         \caption{\small #2\label{#1}}
         \end{figure}}
\newcommand{\comment}[1]{}

\newcommand{\figref}[1]{Fig.~\protect\ref{#1}}

\def\bbbz{{\sf Z\!\!\!Z}}
\newcommand{\PP}{\mbox{I}\!\mbox{P}}
\newcommand{\bbbc}{\mbox{C}\!\!\!\mbox{I}}
\def\sl2z{SL(2,\bbbz)}

\newcommand{\be}{\begin{equation}}
\newcommand{\ee}{\end{equation}}
\newcommand{\bea}{\begin{eqnarray}}
\newcommand{\eea}{\end{eqnarray}}
\newcommand{\nn}{\nonumber}

\newcommand{\inter}[2]{\null^{\#}(#1\cdot#2)}

\preprint{MIT-CTP-3158\\UTTG-06-01\\hep-th/0108137\\}
\title{Quiver Theories from D6-branes via Mirror Symmetry}
\author{Amihay Hanany$^{~1}$\,
and Amer Iqbal$^{~2}$\\

$^{1}$ Center for Theoretical Physics,\\ Massachusetts Institute
of Technology,\\ Cambridge, MA, 02139.\\ \email{hanany@mit.edu}
\vspace{.1in}

$^{2}$ Theory Group, Department of Physics,\\ University of Texas
at Austin,\\ Austin, TX, 78712.\\
\email{iqbal@physics.utexas.edu}}

\abstract{We study ${\cal N}=1$ four dimensional quiver theories
arising on the worldvolume of D3-branes at del Pezzo singularities
of Calabi-Yau threefolds. We argue that under local mirror
symmetry D3-branes become D6-branes wrapped on a three torus in
the mirror manifold. The type IIB $(p,q)$ 5-brane web description
of the local del Pezzo, being closely related to the geometry of
its mirror manifold, encodes the geometry of 3-cycles and is used
to obtain gauge groups, quiver diagrams and the charges of the
fractional branes.}

\begin{document}
\pagenumbering{arabic}
\section{Introduction}
\label{intro}

The importance of mirror symmetry in the study of four dimensional
quantum field theories is well established following the solution
of a large class of ${\cal N}=2$ theories using mirror symmetry
\cite{KKV,KMV}. More recently exact expressions for the
superpotentials in some ${\cal N}=1$ theories has been obtained
using Vafa's large N duality and mirror symmetry \cite{CIV}.

In this paper we will study ${\cal N}=1$, $D=4$ theories obtained by
placing D3-branes at certain singularities of Calabi-Yau
threefolds. The singularities of the Calabi-Yau threefold we will
consider arise from collapsing del Pezzo surfaces \cite{MS, DKV,
IMS}. The case of toric del Pezzo singularities was studied in detail
in \cite{FHH1,FHH2} and gauge groups and quivers diagrams representing
the matter content and their interactions were obtained using partial
resolution \cite{BGLP} of $\bbbc^{~3}/\bbbz_{3}\times \bbbz_{3}$ and
the so called inverse algorithm which was developed in detail in
\cite{FHH1,FHH2}. The case of non-toric del Pezzos, however, is
difficult to analyze using this method. Also the RR charges of the
fractional branes in these geometries cannot be obtained this way. We
will use local mirror symmetry to solve both these problems. The
geometry of 3-cycles in the mirror manifold will provide us with not
only the gauge groups and the quiver diagrams but also RR charge of
the fractional branes. Fractional branes in these geometries
correspond to bundles on the del Pezzo surfaces, the compact divisor
of the non-compact Calabi-Yau space, and are therefore bound states of
D7-branes, D5-branes and D3-branes. These fractional branes are mirror
to 3-cycles which can become massless as we change the complex
structure of the mirror Calabi-Yau manifold. The map between bundles
on del Pezzos and 3-cycles in the mirror Calabi-Yau that we will use
was determined in \cite{HI,MOY2}.  This map is such that the intersection
number between the 3-cycles in the mirror manifold is equal to the
number of fermionic zero modes of strings stretched between the
corresponding fractional branes. In terms of gauge theory data, each
3-cycle maps to a single gauge group factor and the intersection
numbers count the number of ${\cal N}=1$ chiral multiplets which
transform in the bi-fundamental representation of the gauge groups
associated to the two 3-cycles involved. The sign of the intersection
number determines the chirality of the multiplet.

We will see that under local mirror symmetry a 0-cycle of the local
del pezzo surface maps to a three torus in the mirror
manifold. Therefore a D3-brane transverse to the local del Pezzo
becomes a D6-brane wrapping a $T^{3}$ in the mirror manifold. Thus
${\cal N}=1$ D=4 quiver theories that we are interested in can be
obtained from D6-branes wrapping 3-cycles in the mirror manifold. Also
this mirror description involving D6-branes on 3-cycles allows us to
interpret ``toric duality'' \cite{FHH1} as Picard-Lefshetz monodromy
action on the 3-cycles.  This will be discussed in detail elsewhere
\cite{HI2}. For other dualities of ${\cal N}=1$ theories derived from
engineering the theory by D6-branes see \cite{papers}. The resolution
of singularities of Calabi-Yau twofolds and threefolds using
non-commutative algebras was recently studied in \cite{BL} which might
be interesting in D6-brane context for theories related by
Picard-Lefshetz transformations.

The paper is organized as follows. In section \ref{delpezzo} we review
the CY 3-folds mirror to local del Pezzo surfaces and their relation
with affine $E_{N}$ backgrounds. In section \ref{T3} we argue that the
quiver theories arising on the D3-branes at del Pezzo singularities
can also be obtained by wrapping D6-branes on a degenerate $T^{3}$ in
the mirror geometry. The adjacency matrices obtained in these cases
are naturally antisymmetric, being identified with the intersection
matrix of the 3-cycles.  There are, however, quiver theories for which
the adjacency matrix is symmetric. As an example of such a quiver we
review the case of the blown-up conifold in subsection
\ref{symmetric}. In section \ref{localtoric} we review the
construction of manifolds mirror to local toric del Pezzos from the
toric data and show they are the same as
the manifolds given in section \ref{delpezzo}. In this section we also
determine the intersection number of 3-cycles in the mirror
manifold. The intersection matrix determines the quiver diagram of the
corresponding theory. In this section we also give the charges of the
corresponding fractional branes. In section \ref{nontoric} we consider
the case of local non-toric del Pezzos and determine the quiver
diagrams of the corresponding theories from the intersection matrix of
3-cycles in the mirror manifold. RR charges of fractional branes in
these geometries are also given. In the appendix we explain how
fractional brane charges can be calculated using local ${\cal B}_{1}$
($\PP^{2}$ blown up at one point) as an example.

\section{Local del Pezzos and mirror symmetry}
\label{delpezzo}

In this section we review the construction of non-compact Calabi-Yau
threefolds which are mirror to local del Pezzo surfaces\footnote{By
local del Pezzo surface we mean the Calabi-Yau threefold which is the
total space of the anticanonical line bundle over the del Pezzo
surface. Such a non-compact CY threefold can be obtained from a
compact CY which is an elliptic fibration over the del Pezzo surface
by sending the K\"ahler class of the elliptic fiber to infinity
\cite{MOY2}.}. Also we show how the existence of affine $E_{N}$
algebra on both sides allows us to identify 0-cycle with a three torus
in the mirror manifold \cite{HI,MOY2}. We also review the relation
between affine $E_{N}$ backgrounds and 5-brane webs which will be
useful in determining the quiver diagrams.

\subsection{Del Pezzo surfaces and affine $E_{N}$ backgrounds}
\label{affine}

A del Pezzo surface is a two complex dimensional compact surface
with ample canonical class. These surfaces can be obtained either
by  blowing up $0\leq N \leq 8$ points on $\PP^{2}$, ${\cal B}_{N}$ or by
blowing up $0\leq M \leq 7$ points on $\PP^{1}\times \PP^{1},$
$\tilde{{\cal B}}_{M}$ . Not all of these surfaces are different and actually
it turns out that ${\cal B}_{N+1}=\tilde{{\cal B}}_{N}$ for $N\geq 1$.

The basis of $H_{2}({\cal B}_{N},\bbbz)$ is $\{l,E_{1},\cdots, E_{N}\}$,
where $l$ is the pull back of the generator of
$H_{2}(\PP^{2},\bbbz)$ under the projection $\pi:{\cal B}_{N}\mapsto
\PP^{2}$. And $E_{i}$ is the class of the exceptional curve
obtained by blowing up the i-th point on $\PP^{2}$. The
intersection form in this basis is diagonal given by \bea 
\inter{l}
{l}=1\,,\,\inter{l}{E_{i}}=0\,,\,\,\inter{E_{i}}{E_{j}}=-\delta_{ij}\,,\,\,i,j=1,\cdots,N\,. \eea The interesting
property of $H_{2}({\cal B}_{N})$ is that it contains a codimension one
sublattice which is isomorphic to the root lattice of the $E_{N}$
algebra. The simple roots of the algebra are given by curves of
self intersection $-2$ which are orthogonal to the anticanonical class
$K_{B_{N}}=-3l+\sum_{i=1}^{N}E_{i}$, \bea
\alpha_{i}&=&E_{i}-E_{i+1}\,,\,\,i=1,\cdots, N-1\,,\\ \nonumber
\alpha_{N}&=&l-E_{1}-E_{2}-E_{3}\,,\\ \nonumber
\inter{\alpha_{a}}{\alpha_{b}}&=&-A_{ab}^{E_{N}}\,,\,\,a,b=1,\cdots,N\,.\eea
Where $A_{ab}^{E_{N}}$ is the Cartan matrix of the $E_{N}$ algebra.

As mentioned before the non-compact CY containing a del Pezzo surface
${\cal B}_{N}$ is the total space of the anticanonical bundle over
${\cal B}_{N}$. We will denote such a Calabi-Yau threefold by $X_{N}$ and the
corresponding mirror Calabi-Yau threefold by $Y_{N}$. The mirror
manifold $Y_{N}$ is given by the following equations
\cite{HI,MOY2,YY,HV,HIV}\bea y^{2}&=&x^{3}+f^{(N)}(z)x+g^{(N)}(z)\,,\\
\nonumber uv&=&z\,. \eea Explicit expressions for $f^{(N)}(z)$ and
$g^{(N)}(z)$ can be found in \cite{SZ,YY}. The first equation defines
an elliptic fibration over the z-plane. We will denote this two
complex dimensional manifold by ${\cal E}_{N}$. This elliptic
fibration has $N+3$ degenerate fibers with following $(p,q)$ charge 
\bea
\underbrace{[1,0]\cdots
[1,0]}_{N}\,[2,-1]\,[-1,2]\,[-1,-1]\,
\eea
and has total monodromy
$T^{9-N}=\left({1\;\;\; 9-N\atop 0 \;\;\;\;\;\;\;1}\right)$ around the
degenerate fibers. This monodromy allows the existence of a very
special 2-cycle $\Delta$. This 2-cycle is formed by taking a direct
product of a loop surrounding the position of degenerate fibers in the
z-plane and the $(1,0)$ cycle of the elliptic fibration over the
loop. The lattice of 2-cycles in this elliptic fibration contains a
sublattice which is isomorphic to the $E_{N}$ root lattice. It turns
out that since \bea \inter{\Delta}{\Delta}=\inter{\Delta}{C}=0\,,\,\,\,\forall
C\in H_{2}({\cal E}_{N})\,, \eea $\Delta$ can be thought of as an
imaginary root extending the $E_{N}$ algebra to an affine $E_{N}$
algebra \cite{DHIZ}. We will see in the next section that $\Delta$ can
be used to construct a 3-cycle which is the mirror of 0-cycle on
$X_{N}$.

Actually the same structure of affine $E_{N}$ algebra is present in
$H_{*}({\cal B}_{N})$. To see this consider two vector bundles
$V_{1,2}$ on ${\cal B}_{N}$ such that
$\mbox{ch}(V_{a})=(\mbox{r}_{a},\Sigma_{a},\mbox{ch}(V_{a}))$
\footnote{
In the rest of the paper we will use the same symbol for the 2-form
and its dual 2-cycle. Thus $\inter{\Sigma_{a}}{\Sigma_{b}}\equiv
\int_{{\cal B}_{N}} \Sigma_{a}\wedge \Sigma_{b}.$}. An
inner product on the K-theory group of ${\cal B}_{N}$ is given by, \bea
\chi_{{\cal B}_{N}}(V_{1},V_{2})&=&\int_{{\cal B}_{N}}\mbox{ch}(V_{1})\otimes
\mbox{ch}(V^{*}_{2})\mbox{Td}({\cal B}_{N}) \nonumber\\ \nonumber
&=&\int_{{\cal B}_{N}}\mbox{ch}(V_{1})\wedge\mbox{ch}(V^{*}_{2})\wedge
\mbox{Td}({\cal B}_{N})\\ \nonumber
&=&\int_{X}\mbox{ch}(V_{1})\wedge\mbox{ch}(V_{2})^{\vee}\wedge
\mbox{Td}({\cal B}_{N}),
\eea
where $V_{2}^{*}$ is the dual bundle, $\mbox{Td}({\cal B}_{N})=
1+\frac{1}{2}c_{1}({\cal B}_{N})+\frac{1}{12}(c_{1}({\cal B}_{N})^{2}+
c_{2}({\cal B}_{N}))$
and if $v=\sum_{i=0}^{2}v_{i},\, v_{i}\in H^{2i}(B_{N})$ then
$v^{\vee} \equiv \sum_{i=0}^{2}(-1)^{i}v_{i}$. In terms of 
$(\mbox{r}_{a},\Sigma_{a},\mbox{ch}(V_{a}))$ we get
\begin{eqnarray} 
\chi_{{\cal B}_{N}}(V_{1},V_{2})= \mbox{r}_{1}\mbox{r}_{2}-
\inter{\Sigma_{1}}{\Sigma_{2}}+\mbox{r}_{1}\mbox{ch}_{2}(V_{2})
+\mbox{r}_{2}\mbox{ch}_{2}(V_{1})+
\frac{1}{2}(\mbox{r}_{2}\mbox{d}_{\Sigma_{1}}-
\mbox{r}_{1}\mbox{d}_{\Sigma_{2}}).
\end{eqnarray} 
Where $d_{\Sigma}=-K_{{\cal B}_{N}}\cdot \Sigma$.  From the above equation it
follows that this product reduces to intersection numbers when
considering sheaves with support on curves in ${\cal B}_{N}$. We therefore
define the simple roots $\{R_{1},\cdots, R_{N}\}$ such that
$\mbox{ch}(R_{a})=(0,\alpha_{a},-1)$. Then \bea
\chi_{{\cal B}_{N}}(R_{a},R_{b})=A^{E_{N}}_{ab}\,,\,\,a,b=1,\cdots, N\,. \eea
This implies that the K-theory lattice contains a sublattice which is
isomorphic to the $E_{N}$ root lattice. Now consider the K-theory
element $\nabla$ such that $\mbox{ch}(\nabla)=(0,0,-1)$. It follows
that \bea \chi_{{\cal B}_{N}}(\nabla,\nabla)=\chi_{{\cal B}_{N}}(R_{i},\nabla)=0\,,
\eea Thus since $\nabla$ is orthogonal to all the roots and to itself
we see that it realizes the imaginary root of the $E_{N}$ algebra
extending the $E_{N}$ root lattice to the affine $E_{N}$ root lattice
\cite{HI} . We will see in the next section that $\Delta$ defines a
$T^{3}$ in $Y_{N}$ and is therefore mirror to $\nabla$ which has the
charge of a zero cycle. Thus mirror symmetry maps the zero cycle to
the three torus in $Y_{N}$.

\subsection{${\cal E}_{N}$ and 5-brane webs}
It is known that the M-theory on a local del Pezzo leads to a five
dimensional theory which is dual to the theory on a $(p,q)$ five brane
web in type IIB string theory. For the case of local toric del Pezzos
the 5-brane webs were constructed in \cite{webs}. It was shown in
\cite{LV} how this duality between local del Pezzos and 5-brane webs
follows from the duality between M-theory on a torus and type IIB
string theory on $S^{1}$. By compactifying one of the transverse four
spatial directions one can lift the 5-brane web to an M5-brane wrapped
on a non-compact Riemann surface embedded in $\bbbc^{\times}\times
\bbbc^{\times}$.  Local non-toric del Pezzo surfaces, however, do not have
dual 5-brane description since the corresponding 5-brane webs have
external legs which are either parallel or cross each other ruining
the five dimensional interpretation of the theory.

It was shown in \cite{DHIK} that by adding $(p,q)$ 7-branes to the
picture one can obtain a web picture of the local non-toric del
Pezzos. In this case external legs of the 5-brane web are not allowed
to cross each other by making them end on 7-branes. By compactifying
one of the four spatial transverse directions we can lift the 5-brane
configuration to M-theory. In this case we get an M5-brane wrapped on
a Riemann surface which is embedded in a non-compact Calabi-Yau
twofold. This non-compact Calabi-Yau twofold is exactly the affine
$E_{N}$ background, ${\cal E}_{N}$, we mentioned earlier. 

Thus $(p,q)$ 5-brane webs dual to del Pezzo surfaces provide the
complete information about the degenerate fibers of the ${\cal E}_{N}$
which is used in constructing the mirror of the non-compact Calabi-Yau
threefolds containing del Pezzo surface. In the next section we will
show how the information about the charge of degenerate fibers can be
used to determine the gauge groups and the quiver diagrams. In section
4 we will write down the charge of the degenerate fibers directly from
the toric diagram to determine the quivers.

\section{D6-branes on $T^{3}$}
\label{T3}

Low energy Type IIA string theory on a noncompact Calabi-Yau
threefold leads to a four dimensional quantum field theory with
${\cal N}=2$ supersymmetry in the transverse space. The
supersymmetry can be broken down to ${\cal N}=1$ by introducing
D-branes wrapped on appropriate cycles.

D6-branes play an important role in such a construction of ${\cal
N}=1$ theories from Type IIA strings. D6-branes wrapped on special
Lagrangian 3-cycles in the Calabi-Yau threefold preserve ${\cal
N}=1$ supersymmetry on their worldvolume. D6-branes wrapped on
different 3-cycles lead to supersymmetric gauge theories with
matter. The matter content of such a gauge theory is encoded in a
quiver diagram.

In the rest of this section we will restrict ourselves to the
Calabi-Yau manifolds $Y_{N}$ defined in the last section and a
reducible 3-cycle which is topologically a $T^{3}$ constructed
from $\Delta$. As discussed in the previous section the Calabi-Yau
manifolds $Y_{N}$ are defined by the following equations, \bea
uv&=&z\,,\\ \nonumber y^{2}&=&x^{3}+f^{(N)}(z)x+g^{(N)}(z)\,. \eea
The first equation defines a $\bbbc^{\,\times}$ fibration over the
$z$-plane which degenerates at $z=0$. In the second equation
$f^{(N)}(z)$ and $g^{(N)}(z)$ are such that the Weierstrass form defines a
noncompact Calabi-Yau twofold with intersection
matrix of closed two cycles equal to the affine $E_{N}$ Cartan
matrix. The zero cycle of the local del Pezzo is mapped to a
$T^{3}$ in this geometry under mirror symmetry as discussed in
detail in the previous section \cite{HI,MOY2}. We identified these
objects as being mirror to each other because they both represent
the imaginary root of the affine $E_{N}$ algebra in their
respective geometries. This implies that a D3-brane transverse to
a local del Pezzo (so that it is a zero cycle as far as the
Calabi-Yau is concerned) becomes a D6-brane wrapping the mirror
$T^{3}$. In the next section we will show that this is consistent
with the known results for the quiver theory obtained from toric
del Pezzo singularities.

The manifolds ${\cal E}_{N}$ have been studied before in the
context of F-theory and the topology of open and closed curves in
these manifolds is well understood \cite{affine,DHIZ,weyl}. Since 
these backgrounds have monodromy $T^{9-N}\in \sl2z$ there 
exists a two torus formed by taking a closed circle containing all
the points in the $z$-plane over which degenerate fibers are
present and $(1,0)$ cycle of the elliptic fiber as shown in
\figref{twotorus}. This is the curve $\Delta$ mentioned in the
previous section. The $T^{3}$ we are interested in is formed by
taking the two torus in the affine $E_{N}$ background and the
circle of the $\bbbc^{\,\times}$ fibration.

\onefigure{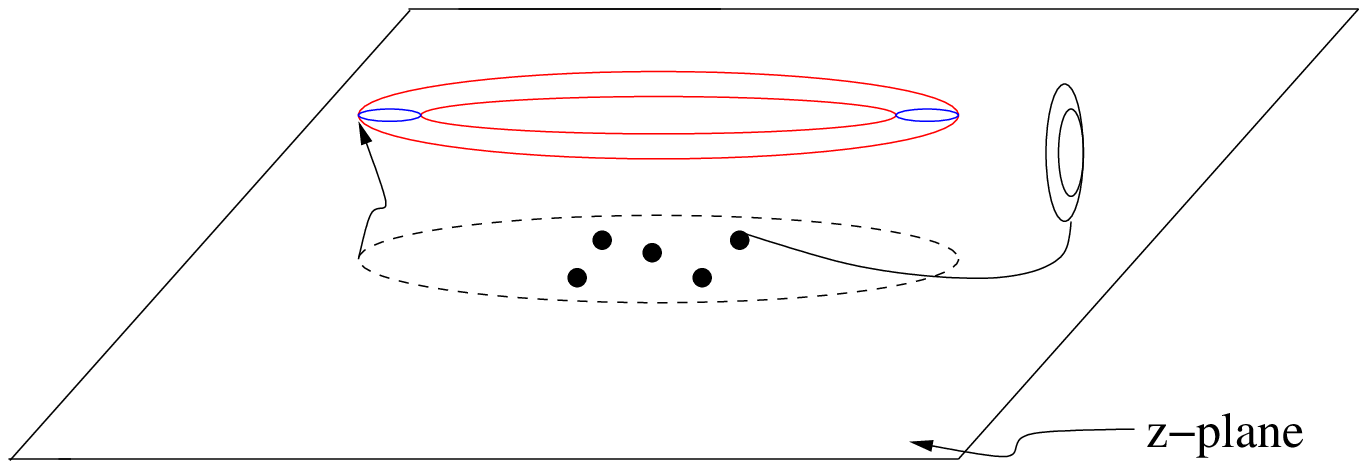}{The cycle $\Delta$ formed by a closed loop in
the base and a 1-cycle in the fiber.}

The homology class of this $T^{3}$ is equal to the sum of the
homology classes of 3-cycles $S_{i}$ which are topologically
$S^{3}$ \cite{DHIZ,HI}. Let us denote by $z_{i}$ the points in the
z-plane where the elliptic fibration degenerates. There are $N+3$
such points for the case of the affine $E_{N}$ background. The
3-cycles $S_{i}$ are formed by taking a path connecting $z_{i}$ to
$z=0$, the 1-cycle of the elliptic fibration which degenerates at
$z_{i}$ and the circle of the $\bbbc^{\,\times}$ fibration. The
geometry of the cycle $S_{i}$ is shown in \figref{s3}. Thus there
are $N+3$ independent 3-cycles in this geometry.

\onefigure{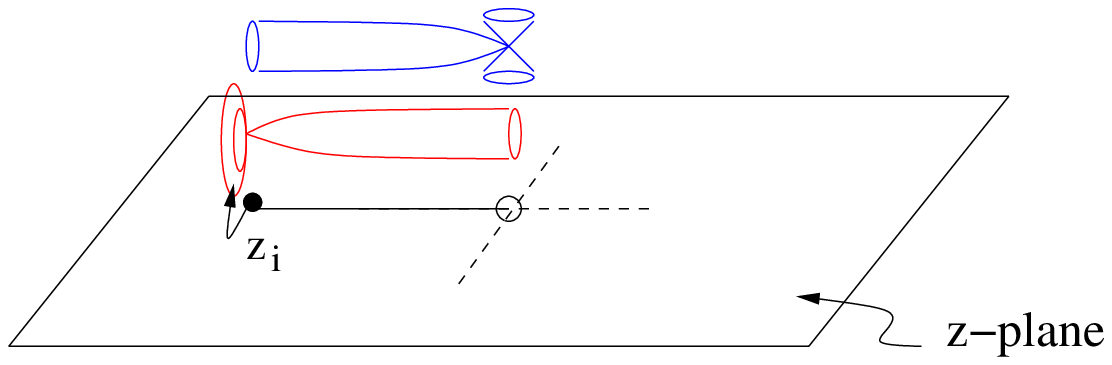}{The 3-cycle with topology of $S^{3}$.}

The intersection number of these 3-cycles can be calculated
easily. From \figref{s3} it is clear that two cycles $S_{i}$ and
$S_{j}$ only intersect above the point $z=0$ as long as $z_{i}\neq
z_{j}$. Above $z=0$ we have a smooth elliptic fiber and the
3-cycle $S_{i}$ wraps a 1-cycle $C_{i}$ of the this elliptic fiber
which degenerates at $z=z_{i}$.  If $C_{i}$ and $C_{j}$ have the
charge $(p_{i},q_{i})$ and $(p_{j},q_{j})$ then the intersection
number, $^{\#}(S_{i}\cdot S_{j})$, is given by
\footnote{$C_{i}=p_{i}\alpha+q_{i}\beta\,,\,\,\alpha,\beta\in
H_{1}(T^{2},\bbbz)$ such that $^{\#}(\alpha\cdot \beta) =1$.}\bea
^{\#}(S_{i}\cdot S_{j})=^{\#}(C_{i}\cdot C_{j})=\mbox{det}
\pmatrix{p_{1} & p_{2}\cr q_{1} & q_{2}}\,. \label{detpq}\eea

Each 3-cycle $S_{i}$ is mirror to a coherent sheaf $F_{i}$ on the
del Pezzo surface ${\cal B}_{N}$. If $f: {\cal B}_{N} \mapsto Y_{N}$ is the
embedding of the del Pezzo surface in Calabi-Yau threefold $Y_{N}$
then we denote by $f_{!}F_{i}$ the coherent sheaf on $Y_{N}$
obtained by extending $F_{i}$ by zero outside ${\cal B}_{N}$ \cite{MOY2}.
A bilinear product which counts the number of fermionic zero modes
of the string stretched between two sheaves $f_{!}F_{i}$ and
$f_{!}F_{j}$ is given by \cite{MOY2}\bea I_{Y_{N}}
=\int_{Y_{N}}\mbox{ch}(f_{!}F_{i})^{*}\mbox{ch}(f_{!}F_{j})\mbox{Td}(Y_{N})
\eea This is an antisymmetric product such that \bea
I_{Y_{N}}(f_{!}F_{i},f_{!}F_{j})=\chi_{{\cal B}_{N}}(F_{i},F_{j})-\chi_{{\cal B}_{N}}(F_{j},F_{i})=
\mbox{det}\pmatrix{r_{1} & r_{2} \cr d_{\Sigma_{1}} & d_{\Sigma_{2}}} \,.
\eea Where $\mbox{ch}(F_{i})=(r_{i},\Sigma_{i},k_{i})$ and
$d_{\Sigma_{i}}=-K_{B_{N}}\cdot \Sigma_{i}$.

Since the map between curves (with and without boundary) in the ${\cal
E}_{N}$ manifold and bundles on del Pezzos given in \cite{HI,MOY2} was
such that \bea r_{i}=q_{i}\,,\\ \nonumber d_{\Sigma_{i}}=p_{i}\,.
\label{mapp} \eea We see that
\bea I_{Y_{N}}(f_{!}F_{i},f_{!}F_{j})=\mbox{det}\pmatrix{r_{1} & r_{2}
\cr d_{\Sigma_{1}} & d_{\Sigma_{2}}} =-\mbox{det} \pmatrix{p_{1} &
p_{2}\cr q_{1} & q_{2}}=-^{\#}(S_{i}\cdot S_{j})\,. \eea Thus the
three cycle on which D6-brane is wrapped is given by \bea
[T^{3}]=\sum_{i=1}^{k+3}S_{a}\,.  \eea and therefore the gauge group $G$
and the quiver matrix $Q_{ij}$ from which quiver diagrams can be constructed is
given by \bea
G&:=&\prod_{i=1}^{k+3}U(1)\,,\\\nn 
Q_{ij}&:=&I_{Y_{N}}(\pi_{!}F_{i},\pi_{!}F_{j})=-^{\#}(S_{i}\cdot
S_{j})\,. \eea 

%
%

\subsection{Symmetric adjacency matrix: The case of the conifold}
\label{symmetric}

In quiver theories the adjacency matrix need not be antisymmetric.
It turns out to be a special feature of toric del Pezzo
singularities. For general singularities there is no principle
which will restrict the adjacency matrix to be antisymmetric and
typically it will have a symmetric as well as an antisymmetric
contribution. Quiver theories which are non-chiral like ${\cal N}=2$
supersymmetric theories have an adjacency matrix which is
symmetric. Many other examples have this generic feature. In this
section we consider such a case, which is relatively simple to
calculate using mirror symmetry, and generates a symmetric
adjacency matrix.

The mirror of the blown-up conifold is given by \cite{vafa}, \bea
(x_{1}+1)(x_{2}+1)-uv=1-e^{-t}\,. \eea This equation for the
mirror manifold can be obtained from the superpotential of the
mirror Landau-Ginzburg theory. The superpotential derived from the
linear sigma model charges $(1,1,-1,-1)$ is \cite{HV}\bea
W(x)=x_{0}+x_{1}+x_{2}+e^{-t}\frac{x_{1}x_{2}}{x_{0}}\,, \eea
where $t$ is the complexified K\"ahler parameter which determines
the size of $\PP^{1}$ in the blownup geometry. The periods in the
LG theory are given by \bea \Pi&:=&\int
e^{-W(x)}\frac{dx_{0}dx_{1}dx_{2}}{x_{0}x_{1}x_{2}} \,\\ &=&\int
e^{-x_{0}(1+\hat{x}_{1}+\hat{x}_{2}+e^{-t}\hat{x}_{1}\hat{x}_{2})}
\frac{dx_{0}d\hat{x}_{1}d\hat{x}_{2}}{x_{0}\hat{x}_{1}\hat{x}_{2}}\,,\eea
where $\hat{x}_{i}=\frac{x_{i}}{x_{0}}$. To be able to integrate
over $x_{0}$ we introduce two more variables $u,v$. \bea
\Pi&=&\int
e^{-x_{0}(1+\hat{x}_{1}+\hat{x}_{2}+e^{-t}\hat{x}_{1}\hat{x}_{2}-uv)}dudv
dx_{0}\frac{d\hat{x}_{1}d\hat{x}_{2}}{\hat{x}_{1}\hat{x}_{2}}\,\\
&=&\int
\delta(1+\hat{x}_{1}+\hat{x}_{2}+e^{-t}\hat{x}_{1}\hat{x}_{2}-uv)
dudv\frac{d\hat{x}_{1}d\hat{x}_{2}}{\hat{x}_{1}\hat{x}_{2}}\,.
\eea Thus the LG periods are equal to the periods of the
holomorphic 3-form \bea \Omega=\frac{dudv
d\hat{x}_{1}d\hat{x}_{2}}{df\hat{x}_{1}\hat{x}_{2}}\,, \eea where
\bea
f:=1+\hat{x}_{1}+\hat{x}_{2}+e^{-t}\hat{x}_{1}\hat{x}_{2}-uv=0\,,
\eea is the equation of the mirror manifold. By rescaling the
variables we can write this equation as \bea
(\hat{x}_{1}+1)(\hat{x}_{2}+1)-uv=1-e^{-t} \eea To understand the
geometry of 3-cycles in this manifold we introduce the variable
$z$ such that \bea (x_{1}+1)(x_{2}+1)=z\,,\\ uv+1-e^{-t}=z\,. \eea
The first equation defines a $\bbbc^{\,\times}$ over the $z$-plane
which degenerates at $z=0$.  The second equation defines another
$\bbbc^{\,\times}$ fibration over the $z$-plane which degenerates
at $z_{*}:=1-e^{-t}$. As discussed before, these two fibrations
can be used to define an $S^{3}$ which shrinks as $t\mapsto 0$.
There also exists a second 3-cycle in this geometry which is
topologically a $T^{3}$. This cycle is formed by taking a closed
loop encircling the points $z=0$ and $z=z_{*}$ together with the
circles of the two $\bbbc^{\,\times}$ fibrations.  This $T^{3}$ is
actually the sum of two $S^{3}$ as shown in \figref{conifold}
below. \onefigure{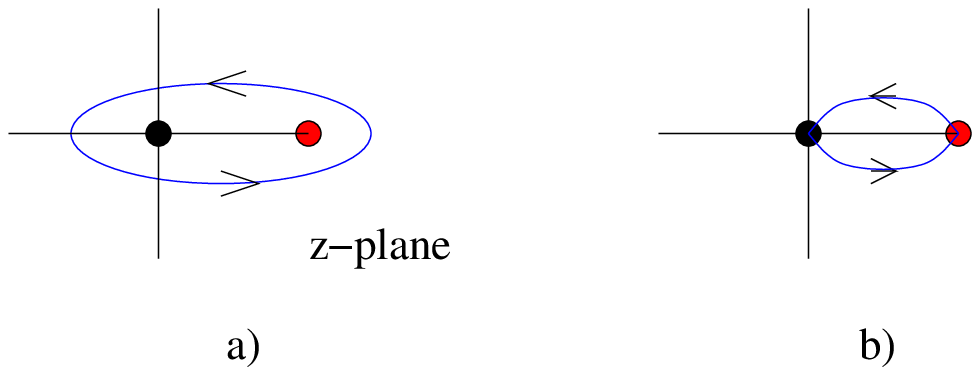}{3-cycles in the mirror of the conifold.}

The two $S^{3}$'s intersect each other at two points one above
$z=0$ and the other one above $z=z_{*}$. Since the
self-intersection number of $[T^{3}]$ is zero there are two
hypermultiplets in the representation $(N,\bar{N})$ and
$(\bar{N},N)$ in the corresponding gauge theory. The quiver
diagram of the gauge theory obtained by wrapping D6-branes on this
$T^{3}$ is shown in \figref{conifold-quiver} below.
\onefigure{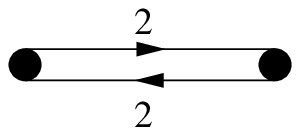}{Quiver diagram of the gauge theory on
the D3-brane transverse to a conifold singularity.}

\section{Local toric del Pezzos}
\label{localtoric}

The geometry of the local toric del Pezzos is completely determined by
the corresponding del Pezzo surface. These Calabi-Yau manifolds are
the total space of the anticanonical bundle over the del Pezzo
surface. The toric data of these Calabi-Yau manifolds is encoded in
the diagrams shown in \figref{toric} below.

\onefigure{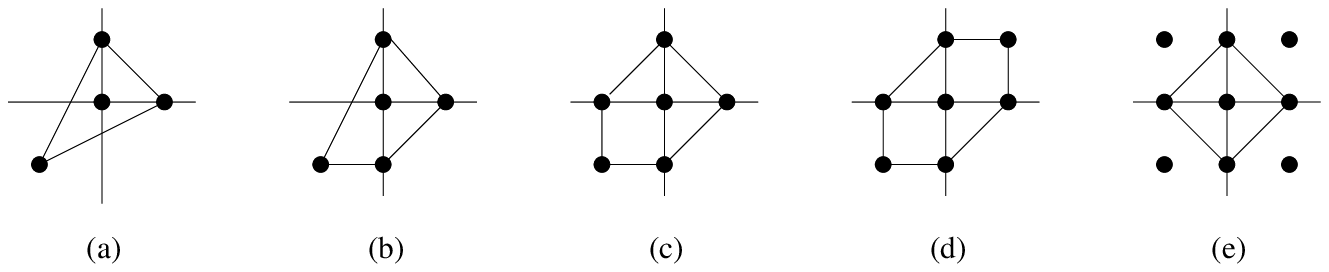}{Toric diagrams for the del Pezzo surfaces. a)
$\PP^{2}$, b) $\PP^{2}$ blown up at one point, c) $\PP^{2}$ blown
up at two points, d) $\PP^{2}$ blown up at three points, e)
$\PP^{1}\times \PP^{1}$.}

As shown in \cite{LV}, once the toric data is known it is easy to
write down the mirror manifolds, \bea a_{0}+\sum_{\vec{v}\in V}
a_{\vec{v}}x_{1}^{v_{1}}x_{2}^{v_{2}}=uv\,. \eea In the above
equation $V$ is the set of vertices of the diagram given in
\figref{toric} and $v_{i}$ are the coordinates of the vertex and
the variables $x_{i}$ are $\bbbc^{\, \times}$ variables. This
equation for the mirror manifold can also be obtained from the
superpotential of the mirror Landau-Ginzburg theory following the
steps similar to the case of the conifold in the previous section
\cite{HV,HIV}.

\subsection{Geometry of the mirror manifold}

We consider the case of Calabi-Yau obtained from $\PP^{2}$ blown
up at three points. Other local toric del Pezzos can be obtained
from this by blowing down exceptional curves. From \figref{toric}
it follows that the mirror manifold is given by \bea
a_{0}+a_{1}x_{1}+a_{2}x_{2}+\frac{a_{3}}{x_{1}}+\frac{a_{4}}{x_{2}}+
a_{5}\frac{x_{1}}{x_{2}}+a_{6}\frac{x_{2}}{x_{1}}=uv\,. \eea Since
$x_{i}$ are $\bbbc^{\,\times}$ variables we can rescale them and
simplify the above equation, \bea
1+x_{1}+x_{2}+\frac{e^{-t_{1}}}{x_{1}}+\frac{e^{-t_{2}}}{x_{2}}+
e^{-t_{3}}\frac{x_{1}}{x_{2}}+e^{-t_{4}}\frac{x_{2}}{x_{1}}=uv\,.
\eea By introducing another complex variable $w$ we can write the
above equation as a cubic polynomial in $\PP^{2}$ and thus
representing a genus one curve, \bea \nonumber
z\,x_{1}x_{2}w&=&x_{1}x_{2}(x_{1}+x_{2}+w)+wx_{2}(e^{-t_{1}}w+
e^{-t_{4}}x_{2})+wx_{1}(e^{-t_{2}}w+e^{-t_{3}}x_{1})\,,\\
z&=&uv\,. \eea The complex structure parameters $a_{i}$ of the
mirror manifold are related to the K\"ahler structure parameters
$\{t_{1},t_{2},t_{3},t_{4}\}$ of the original manifold. By
redefining coordinates we can write the first equation in the
Weierstrass form representing the manifold ${\cal E}_{3}$. The
second equation defines a $\bbbc^{\,\times}$ fibration over the
complex plane with coordinate $z$. The circle of this fibration
degenerates at $z=0$. The first equation defines an elliptic
fibration over the $z$-plane. The fibration degenerates at six
points on the $z$-plane, $\{z_{*,i}~|~i=1,\cdots, 6\}$. The
position of the degenerate fibers depend on the K\"ahler
parameters $\{t_{1},t_{2},t_{3},t_{4}\}$.

As discussed in the previous section. Using these two fibrations
we can construct 3-cycles with topology of an $S^{3}$ such that
their sum is topologically a $T^{3}$. The matter content of the
gauge theory obtained by wrapping D6-branes on this $T^{3}$ is
encoded in the intersection matrix of the basis 3-cycles $S_{i}$.
In order to calculate the intersection matrix we need the charges
of the 1-cycles of the elliptic fibration which degenerate at
$z=z_{i}$. These charges are not hard to determine since
degenerate fibers of the ${\cal E}_{N}$ are already classified and
correspond to the charges of the external legs of the toric
diagram of local toric del Pezzos \cite{DHIK}. Thus the
intersection number of 3-cycles in the mirror geometry is given by
the toric diagram of the original manifold.

The fractional branes in the original geometry are mirror to
3-cycles which can become massless as we change the complex
structure of the mirror manifold. The cycles $S_{i}$ that we
defined earlier from the basis of such massless 3-cycles and are
mirror to the fractional branes. The fractional branes have the RR
charge of an exceptional bundle on the del Pezzo surface.  This is
because on the mirror side $b_{1}(S^{3})=0$, implying that the
moduli space of $S^{3}$ is zero dimensional. In \cite{HI} the map
between bundles on del Pezzo surfaces and 3-cycles in the mirror
geometry was given.  An important point to keep in mind is the
Picard Lefshetz monodromy action on the 3-cycles in the mirror
geometry which implies that gauge theories with different matter
content can be obtained from the same underlying geometry. This
phenomenon was termed ``toric duality'' in \cite{FHH1}. Different
quiver theories coming from the same del Pezzo geometries were
listed in that paper with a larger set of examples given in
\cite{FHH2}. We will discuss this phenomenon in detail elsewhere
\cite{HI2}.

\underline{\bf local ${\cal B}_{0}$:}

This case has been studied in several papers \cite{DG,DFR,MOY1}
and can be treated by the usual orbifold methods since it is the
resolution of $\bbbc^{\,3}/\bbbz_{3}$. However, we include this
case here for completeness.  The charge of the vanishing cycles of
the elliptic fibration over the $z$-plane, which is part of the
mirror manifold, is determined by the toric diagram of local
$\PP^{2}$, \figref{toricb0}.  \onefigure{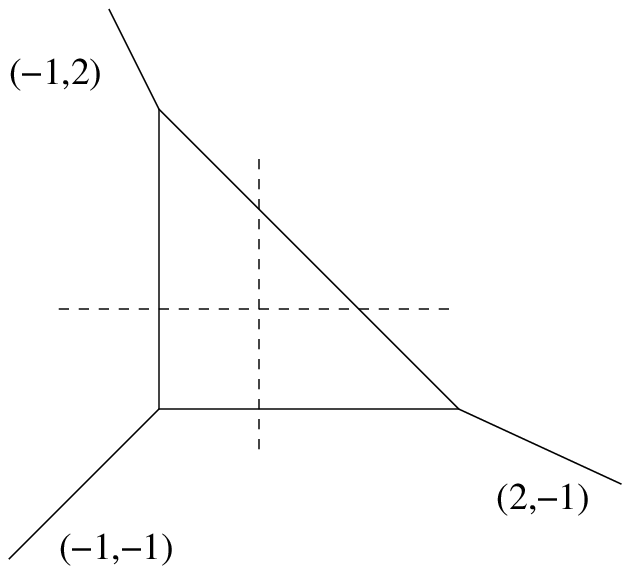}{The toric
diagram of local $\PP^{2}$.} This diagram and the diagrams which
follow coincide with the $(p,q)$ 5-brane webs of five dimensional
field theories \cite{webs}.

The vanishing 1-cycles which define the 3-cycles in the mirror
geometry are \bea C_{1}=(2,-1)
\,,\,\,\,\,C_{2}=(-1,2)\,,\,\,\,\,C_{3}=(-1,-1)\,, \eea and, using
equation (\ref{detpq}), the intersection matrix of corresponding
3-cycles $S_{i}$ is \bea {\cal I}^{0}_{ij}:=^{\#}(S_{i}\cdot
S_{j})=^{\#}(C_{i}\cdot C_{j})=\pmatrix{0 & 3 & -3 \cr -3 & 0 &
3\cr 3 & -3 &0}\,.  \eea Since there are no mutually local
1-cycles the gauge group is abelian and is just \bea
G=\mbox{U}(1)\times \mbox{U}(1)\times \mbox{U}(1)\,. \eea

>From the above intersection matrix we obtain the quiver diagram of
the gauge theory on the D3-branes transverse to the singularity
$\bbbc^{\,3}/\bbbz_{3}$ \figref{quiver-b0}.
\onefigure{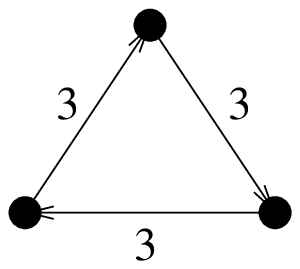}{The quiver diagram obtained from the
intersection matrix ${\cal I}^{0}_{ij}$. The integer on the line
indicates the multiplicity of chiral multiplets.}

There are three fractional branes, which we will denote by
$F_{1,2,3}$, in this geometry mirror to $S_{1,2,3}$. They are
bound states of D7, D5 and D3-branes wrapped on the 4-cycle and
various 2-cycles of the del Pezzo. These fractional branes in the
large volume limit can be identified with vector bundles on the
del Pezzo surface. As shown in the appendix in this geometry they
have the following charges \footnote{For a vector bundle $V$,
$\mbox{ch}(V)$ denotes the Chern character of $V$ and
$k=\int_{{\cal B}_{0}} \mbox{ch}_{2}$. Same notation is going to
be used throughout the paper.}, \bea
S_{1}\,:\,\,\mbox{ch}(F_{1})&=&(-1, l, -\frac{1}{2})\,,\\
\nonumber S_{2}\,:\,\,\mbox{ch}(F_{2})&=&(2,-l,-\frac{1}{2})\,,\\
\nonumber S_{3}\,\,:\,\,\mbox{ch}(F_{3})&=&(-1,0,0)\,. \nonumber
\eea Where $l$ is the generator of $H_{2}(\PP^{2},\bbbz)$.

\underline{\bf local ${\cal B}_{1}$:}

In this case we cannot use the usual orbifold techniques directly to
determine the quiver. The toric diagram of the local ${\cal B}_{1}$ is
shown in \figref{toricb1} below.  \onefigure{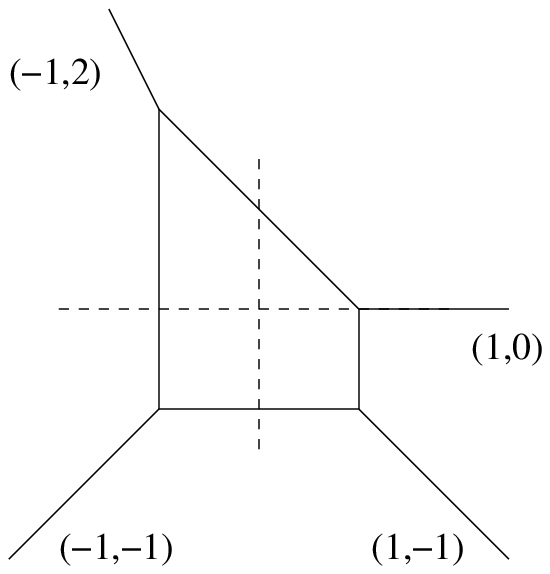}{The toric
diagram of local ${\cal B}_{1}$.}

From the toric diagram it follows that the charges of the
vanishing 1-cycles defining the 3-cycles in the mirror geometry
are given by \bea
C_{1}=(1,-1)\,,\,\,C_{2}=(1,0)\,,\,\,C_{3}=(-1,2)\,,\,\,C_{4}=(-1,-1)\,.
\eea The intersection matrix of the corresponding 3-cycles is
given by \bea {\cal I}^{1}_{ij}:=^{\#}(S_{i}\cdot
S_{j})=^{\#}(C_{i}\cdot C_{j})= \pmatrix{0 & 1& 1&-2\cr
-1&0&2&-1\cr -1&-2&0&3\cr 2&1&-3&0}\,. \eea

As in the case of ${\cal B}_{0}$ we see that there are no mutually
local 1-cycles and therefore the gauge group, $G$, is given by
\bea G=\mbox{U}(1)\times \mbox{U}(1)\times \mbox{U}(1)\times
\mbox{U}(1)\,. \eea The quiver diagram obtained from the above
intersection matrix is shown in \figref{quiver-b1}.
\onefigure{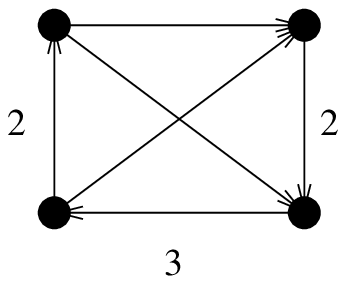}{The quiver diagram obtained from the
intersection matrix ${\cal I}^{1}_{ij}$.}

The fractional branes, denoted by $F_{i}$, in this geometry have
the following charges, \bea
S_{1}\,:\,\,\mbox{ch}(F_{1})&=&(-1,l-E_{1},0)\,,\\ \nonumber
S_{2}\,:\,\,\mbox{ch}(F_{2})&=&(0,E_{1},-\frac{1}{2})\,,\\
\nonumber S_{3}\,:\,\,\mbox{ch}(F_{3})&=&(2,-l,-\frac{1}{2})\,,\\
\nonumber S_{4}\,:\,\,\mbox{ch}(F_{4})&=&(-1,0,0)\,.  \nonumber
\eea Where $E_{1}$ is the exceptional curve which together with
$l$ forms the basis of $H_{2}({\cal B}_{1})$ such that
$^{\#}(E_{1}\cdot E_{1})=-1$ and $^{\#}(l\cdot E_{1})=0$.

\underline{\bf local $F_{0}$:}

The toric diagram of local $F_{0}$ is shown in the
\figref{toricf0} below. \onefigure{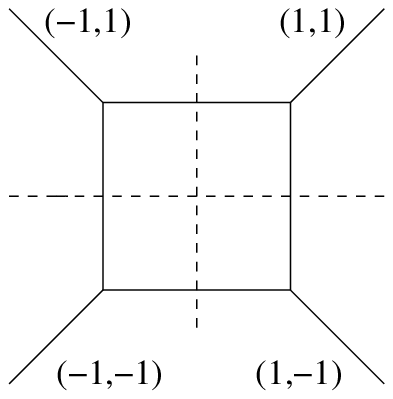}{The toric diagram of
local $\PP^{1}\times \PP^{1}$.} For this case the vanishing
1-cycles are \bea
C_{1}=(1,-1)\,,\,C_{2}=(1,1)\,,\,C_{3}=(-1,1)\,,\,C_{4}=(-1,-1)\,.
\eea The corresponding intersection matrix is \bea I_{ij}=
\pmatrix{0& 2&-2& 0\cr -2 & 0 & 0&2\cr 2&0&0&-2\cr 0&-2&2&0}\,.
\eea And the corresponding quiver diagram is
\onefigure{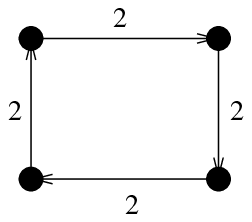}{Quiver diagram obtained from ${\cal
I}_{ij}$.}

The fractional branes have the following charges, \bea
S_{1}\,:\,\,\mbox{ch}(F_{1})&=&(-1,l_{1},0)\,, \\  \nonumber
S_{2}\,:\,\,\mbox{ch}(F_{2})&=&(1,l_{2}-l_{1},-1)\,,\\ \nonumber
S_{3}\,:\,\,\mbox{ch}(F_{3})&=&(1,-l_{2},0)\,,\\  \nonumber
S_{4}\,:\,\,\mbox{ch}(F_{4})&=&(-1,0,0)\,. \eea

\underline{\bf local ${\cal B}_{2}$:}

Now consider the case of local ${\cal B}_{2}$. In this case the
toric diagram is shown in \figref{toricb2} below.
\onefigure{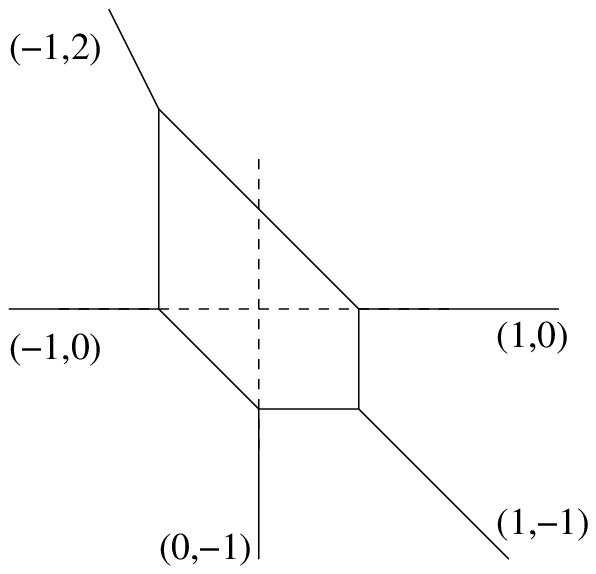}{The toric diagram of local $B_{2}$.}

The vanishing 1-cycles are \bea
C_{1}=(1,-1)\,,\,C_{2}=(1,0)\,,\,C_{3}=(-1,2)\,,\,
C_{4}=(-1,0)\,,\, C_{5}=(0,-1)\,. \eea The intersection matrix of
3-cycles is given by \bea {\cal I}^{2}_{ij}:=^{\#}(S_{i}\cdot
S_{j})=^{\#}(C_{i}\cdot C_{j})= \pmatrix{0&1&1&-1&-1\cr
-1&0&2&0&-1\cr -1&-2&0&2&1\cr1&0&-2&0&1\cr1&1&-1&-1&0}\,.  \eea
The corresponding quiver diagram is shown in \figref{quiver-b2}.
\onefigure{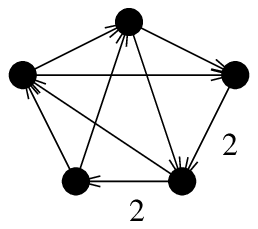}{The quiver diagram obtained from ${\cal
I}^{2}_{ij}$. The integers on the lines indicate multiplicity.}

Since all vanishing 1-cycles are mutually non-local the gauge
group, $G$, is \bea G=\mbox{U}(1)\times \mbox{U}(1)\times
\mbox{U}(1)\times \mbox{U}(1)\times \mbox{U}(1)\,. \eea

The charges of the fractional branes in this geometry are, \bea
S_{1}\,:\,\,\mbox{ch}(F_{1})&=&(-1,l-E_{1},0)\,,\,\,\\ \nonumber
S_{2}\,:\,\,\mbox{ch}(F_{2})&=&(0,E_{1},-\frac{1}{2})\,,\,\,\\
\nonumber S_{3}\,:\,\,\mbox{ch}(F_{3})&=&(2,-l,-\frac{1}{2})\,, \\
\nonumber
S_{4}\,:\,\,\mbox{ch}(F_{4})&=&(0,-E_{2},-\frac{1}{2})\,,\,\,\\
\nonumber S_{5}\,:\,\,\mbox{ch}(F_{5})&=&(-1,E_{2},\frac{1}{2})\,.
\eea

\underline{\bf local ${\cal B}_{3}$:}

${\cal B}_{3}$ is the last toric del Pezzo and is given by
$\PP^{2}$ blownup at three points or $\PP^{1}\times \PP^{1}$
blownup at two points. The toric diagram is shown in
\figref{toricb3}. \onefigure{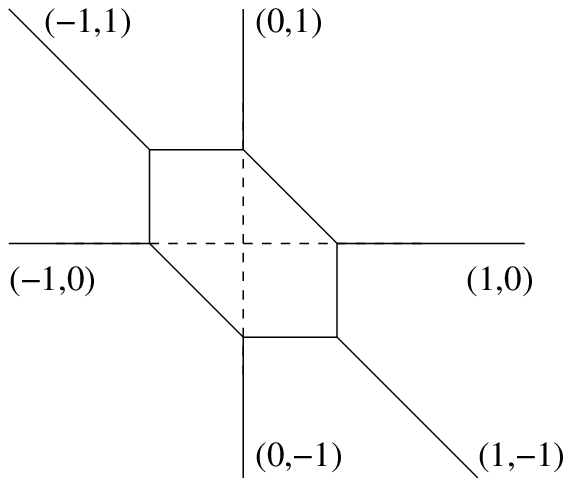}{Toric diagram of local
$B_{3}$}

From the toric diagram we can determine the charge of the
vanishing 1-cycles, \bea
C_{1}&=&(1,-1)\,,\,C_{2}=(1,0)\,,\,C_{3}=(0,1)\,,\, \\ \nonumber
C_{4}&=&(-1,1)\,,\, C_{5}=(-1,0)\,,\,C_{6}=(0,-1)\,.\\ \nonumber
\eea The intersection matrix of corresponding 3-cycles is, \bea
{\cal I}^{3}_{ij}=\pmatrix{0&1&1&0&-1&-1\cr -1&0&1&1&0&-1\cr
-1&-1&0&1&1&0\cr 0&-1&-1&0&1&1\cr 1&0&-1&-1&0&1\cr
1&1&0&-1&-1&0}\,. \eea

The corresponding quiver diagram is shown in \figref{quiver-b3}
below. \onefigure{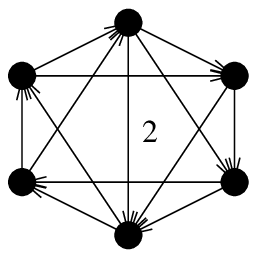}{Quiver diagram obtained from ${\cal
I}^{3}_{ij}$.}

The fractional branes for these cases are \bea
S_{1}\,:\,\,\mbox{ch}(F_{1})&=&(-1,l-E_{1},0)\,,\,\,\\ \nonumber
S_{2}\,:\,\,\mbox{ch}(F_{2})&=&(0,E_{1},-\frac{1}{2})\,,\,\,\\
\nonumber S_{3}\,:\,\,\mbox{ch}(F_{3})&=&(1,-E_{3},-\frac{1}{2})\\
\nonumber S_{4}\,:\,\,\mbox{ch}(F_{4})&=&(1,-l+E_{3},0)\,, \\
\nonumber
S_{5}\,:\,\,\mbox{ch}(F_{5})&=&(0,-E_{2},-\frac{1}{2})\,,\,\,\\
\nonumber S_{6}\,:\,\,\mbox{ch}(F_{6})&=&(-1,E_{2},\frac{1}{2})\,.
\eea

Note that in all the cases discussed above the sum of fractional
branes is equal to a 0-cycle and is required by mirror symmetry.

\section{Non-Toric local del Pezzos}
\label{nontoric}

Non-toric del Pezzo surfaces are obtained by blowing up $N$
($4\leq N\leq 8$) points of $\PP^{2}$. We will denote, as before,
by $E_{i}$ the i-th exceptional curve. It was shown in \cite{DHIK}
that diagrams similar to the toric diagram of the previous section
can also be drawn for non-toric del Pezzos. The important
difference between these and the toric diagrams is that the
diagrams for non-toric del Pezzos will have parallel legs. This is
simply the fact that in the mirror manifold the elliptic fibration
has mutually local vanishing 1-cycles. The elliptic fibration of
the mirror manifold in this case are the affine $E_{N}$ ($4\leq
N\leq 8$) backgrounds \cite{HI}. The charge of the vanishing
1-cycles are known in these cases and therefore we can write down
the quiver diagrams and the gauge groups which will have
non-abelian factors for such cases.

\underline{\bf local ${\cal B}_{4}$:}

We consider cases of $\PP^{2}$ blown up at four points here. The
web picture is shown in \figref{toricb4}. The calculation proceeds
with no essential difficulties.

\onefigure{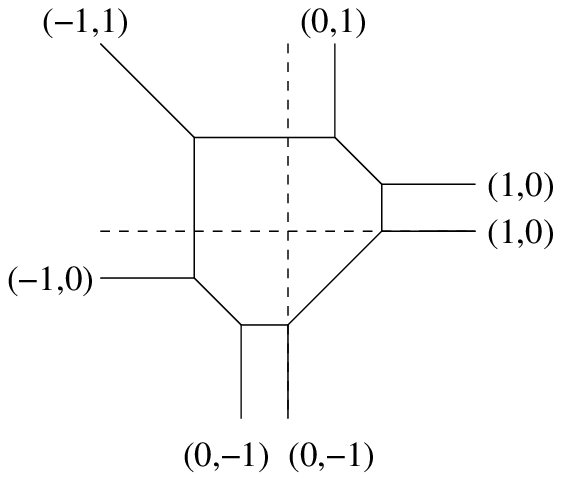}{The web diagram of local $B_{4}$.} We have the
following basis of 1-cycles for the geometry in \figref{toricb4}.
\bea \nonumber C_{1}&=&(0,-1)\,,\,C_{2}=C_{3}=(1,0)\,,\,C_{4}=(0,1)\,,
\\ \nonumber
\,C_{5}&=&(-1,1)\,,\,C_{6}=(-1,0)\,,\,C_{7}=(0,-1)\,. \eea The gauge
group in this case is, \bea G=\prod_{a=1}^{7}\mbox{U}(1)_{a}\,. \eea The
corresponding intersection matrix which determines the quiver diagram
is, \bea I_{ij}=C_{i}\cdot C_{j}=\pmatrix{0&1&1&0&-1&-1&0\cr -1&0&0&1&1&0&-1\cr
-1&0&0&1&1&0&-1\cr 0&-1&-1&0&1&1&0\cr 1&-1&-1&-1&0&1&1\cr
1&0&0&-1&-1&0&1\cr 0&1&1&0&-1&-1&0}\,. \eea The above intersection
matrix gives the quiver diagram shown in \figref{quiver-b4} below.
\onefigure{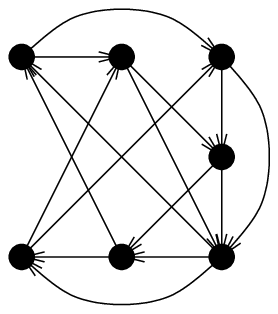}{Quiver diagram obtained from the above
intersection matrix $I_{ij}$.}  The fractional branes for this case
are \bea
S_{1}\,:\,\,\mbox{ch}(F_{1})&=&(-1,l-E_{1}-E_{4},\frac{1}{2})\\
\nonumber
S_{2}\,:\,\,\mbox{ch}(F_{2})&=&(0,E_{4},-\frac{1}{2})\,,\,\,\\
\nonumber
S_{3}\,:\,\,\mbox{ch}(F_{3})&=&(0,E_{1},-\frac{1}{2})\,,\,\,\\
\nonumber S_{4}\,:\,\,\mbox{ch}(F_{4})&=&(1,-E_{3},-\frac{1}{2})\\
\nonumber S_{5}\,:\,\,\mbox{ch}(F_{5})&=&(1,-l+E_{3},0)\\ \nonumber
S_{6}\,:\,\,\mbox{ch}(F_{6})&=&(0,-E_{2},-\frac{1}{2})\,,\,\,\\
\nonumber S_{7}\,:\,\,\mbox{ch}(F_{7})&=&(-1,E_{2},\frac{1}{2})\,.
\eea 


\underline{\bf local ${\cal B}_{5}$}

The web picture for this case is shown in \figref{toricb5} below.
\onefigure{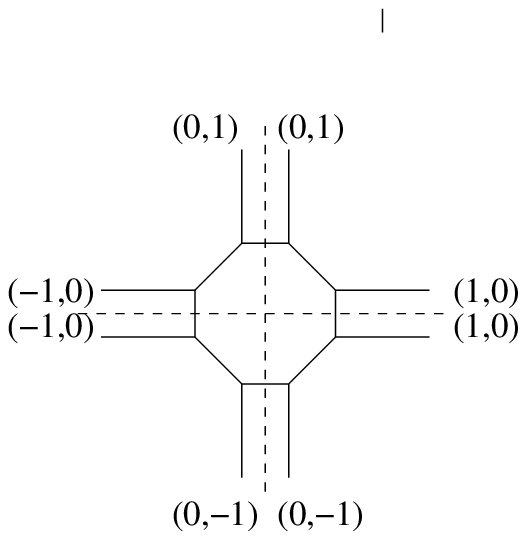}{The web diagram of local $B_{5}$.}

From \figref{toricb5} we have the following 1-cycles, \bea \nonumber
C_{1}&=&(0,-1)\,,\,C_{2}=C_{3}=(1,0)\,,\,C_{4}=C_{5}=(0,1)\,,\,C_{6}=C_{7}=(-1,0)\,,\,C_{8}=(0,-1)\,.
\eea The gauge group, $G$, is given by \bea G=\prod_{a=1}^{8}\mbox{U}(1)_{a}\,. \eea The quiver
diagram corresponding to the intersection matrix is shown in
\figref{quiver-b5}. \onefigure{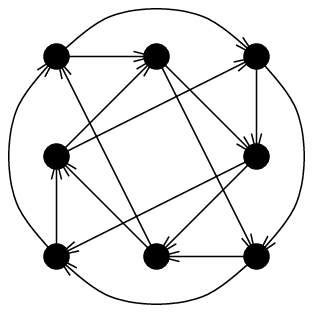}{Quiver diagram obtained from
the intersection matrix of $C_{i}$'s.}

The charge of the fractional branes is given by \bea
S^{a}_{1}\,:\,\,\mbox{ch}(F^{a}_{1})&=&(-1,l-E_{1}-E_{4},\frac{1}{2})\\
\nonumber
S^{a}_{2}\,:\,\,\mbox{ch}(F^{a}_{2})&=&(0,E_{4},-\frac{1}{2})\,,\,\,\\
\nonumber
S^{a}_{3}\,:\,\,\mbox{ch}(F^{a}_{3})&=&(0,E_{1},-\frac{1}{2})\,,\,\,\\
\nonumber
S^{a}_{4}\,:\,\,\mbox{ch}(F^{a}_{4})&=&(1,-E_{3},-\frac{1}{2})\\
\nonumber
S^{a}_{5}\,:\,\,\mbox{ch}(F^{a}_{5})&=&(1,-E_{5},-\frac{1}{2})\\
\nonumber
S^{a}_{6}\,:\,\,\mbox{ch}(F^{a}_{6})&=&(0,-l+E_{3}+E_{5},\frac{1}{2})\\
\nonumber
S^{a}_{7}\,:\,\,\mbox{ch}(F^{a}_{7})&=&(0,-E_{2},-\frac{1}{2})\,,\,\,\\
\nonumber
S^{a}_{8}\,:\,\,\mbox{ch}(F^{a}_{8})&=&(-1,E_{2},\frac{1}{2})\,.
\eea 

\underline{\bf local ${\cal B}_{6}$}

The web picture for this case is shown in \figref{toricb6} below.
For this case and later ones we will only give the charge of the
1-cycles from which the intersection matrix and the quiver diagram
can be obtained easily. \onefigure{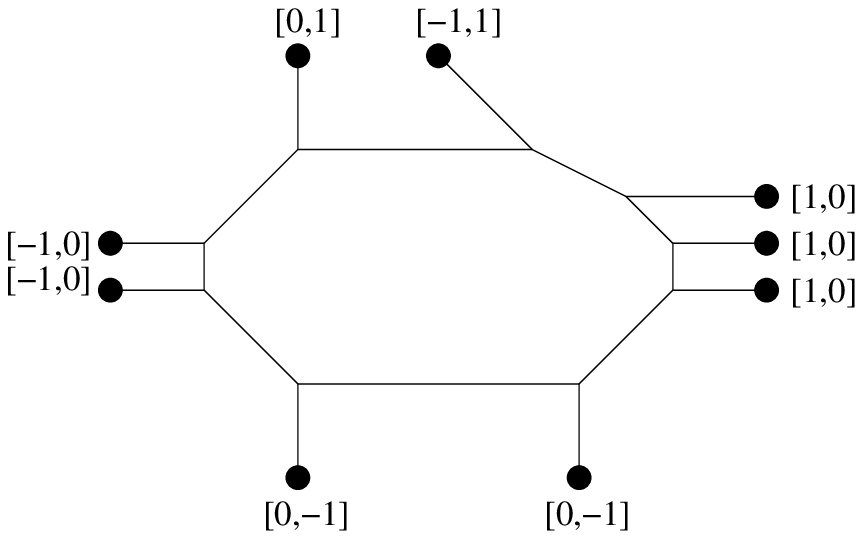}{The web diagram of
local $B_{6}$.}

The 1-cycles are \bea
C_{1}&=&(0,-1)\,,\,C_{2}=C_{3}=C_{4}=(1,0)\,,\,C_{5}=(-1,1)\,,\,C_{6}=(0,1)\,,\,
\\ \nonumber
C_{7}&=&C_{8}=(-1,0)\,,\,C_{9}=(0,-1)\,. \eea The gauge group,
$G$, in this case is, \bea
G=\prod_{a=1}^{9}\mbox{U}(1)_{a}\,. \eea

The fractional brane charges for this geometry are given below,
\bea
S_{1}\,:\,\,\mbox{ch}(F_{1})&=&(-1,l-E_{1}-E_{4},\frac{1}{2})\\
\nonumber
S_{2}\,:\,\,\mbox{ch}(F_{2})&=&(0,E_{4},-\frac{1}{2})\,,\,\,\\
\nonumber
S_{3}\,:\,\,\mbox{ch}(F_{3})&=&(0,E_{1},-\frac{1}{2})\,,\,\,\\
\nonumber
S_{4}\,:\,\,\mbox{ch}(F_{4})&=&(0,l-E_{3}-E_{6},-\frac{1}{2})\\
\nonumber S_{5}\,:\,\,\mbox{ch}(F_{5})&=&(1,-l+E_{6},0)\\
\nonumber S_{6}\,:\,\,\mbox{ch}(F_{6})&=&(1,-E_{5},-\frac{1}{2})\\
\nonumber
S_{7}\,:\,\,\mbox{ch}(F_{7})&=&(0,-l+E_{3}+E_{5},\frac{1}{2})\\
\nonumber
S_{8}\,:\,\,\mbox{ch}(F_{8})&=&(0,-E_{2},-\frac{1}{2})\,,\,\,\\
\nonumber S_{9}\,:\,\,\mbox{ch}(F_{9})&=&(-1,E_{2},\frac{1}{2})\,.
\eea

\underline{\bf local ${\cal B}_{7}$}

The web diagram is shown in \figref{toricb7} below.
\onefigure{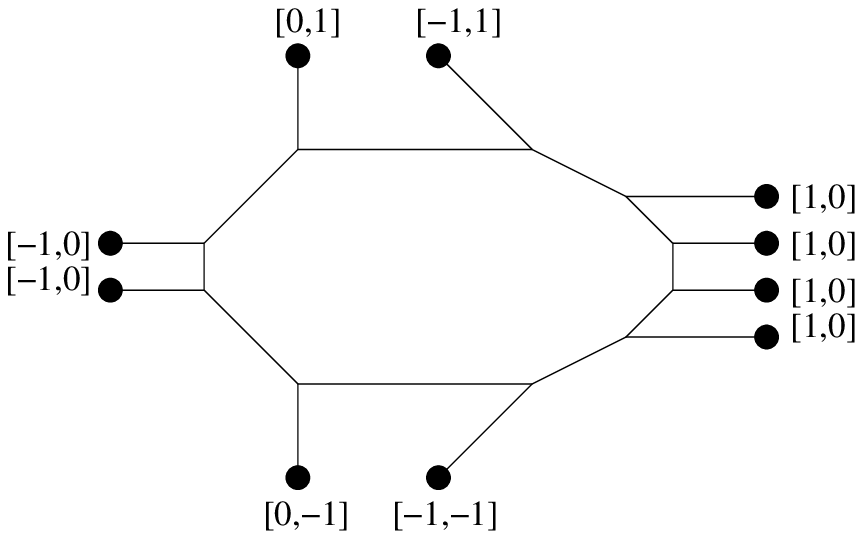}{The web diagram of local $B_{7}$.} The
1-cycles follow from the web diagram \bea
C_{1}&=&(-1,-1)\,,\,C_{2}=C_{3}=C_{4}=C_{5}=(1,0)\,,\,C_{6}=(-1,1)\,,\,
\\ \nonumber
C_{7}&=&(0,1)\,,\,C_{8}=C_{9}=(-1,0)\,,\,C_{10}=(0,-1)\,. \eea The
gauge group in this case is, \bea G=\prod_{a=1}^{10}\mbox{U}(1)_{a}\,. \eea The fractional brane charges
are, \bea
S_{1}\,:\,\,\mbox{ch}(F_{1})&=&(-1,l-E_{1}-E_{4}-E_{7},1)\\
\nonumber S_{2}\,:\,\,\mbox{ch}(F_{2})&=&(0,E_{7},-\frac{1}{2})\\
\nonumber
S_{3}\,:\,\,\mbox{ch}(F_{3})&=&(0,E_{4},-\frac{1}{2})\,,\,\,\\
\nonumber
S_{4}\,:\,\,\mbox{ch}(F_{4})&=&(0,E_{1},-\frac{1}{2})\,,\,\,\\
\nonumber
S_{5}\,:\,\,\mbox{ch}(F_{5})&=&(0,l-E_{3}-E_{6},-\frac{1}{2})\,,\,\\
\nonumber S_{6}\,:\,\,\mbox{ch}(F_{6})&=&(1,-l+E_{6},0)\\
\nonumber S_{7}\,:\,\,\mbox{ch}(F_{7})&=&(1,-E_{5},-\frac{1}{2})\\
\nonumber
S_{8}\,:\,\,\mbox{ch}(F_{8})&=&(0,-l+E_{3}+E_{5},\frac{1}{2})\\
\nonumber S_{9}\,:\,\,\mbox{ch}(F_{9})&=&(0,-E_{2},-\frac{1}{2})\\
\nonumber
S_{10}\,:\,\,\mbox{ch}(F_{10})&=&(-1,E_{2},\frac{1}{2})\,,\,\,\\
\nonumber \eea


\underline{\bf local ${\cal B}_{8}$}

This is the last del Pezzo surface and the corresponding web
diagram is shown in \figref{toricb8}. \onefigure{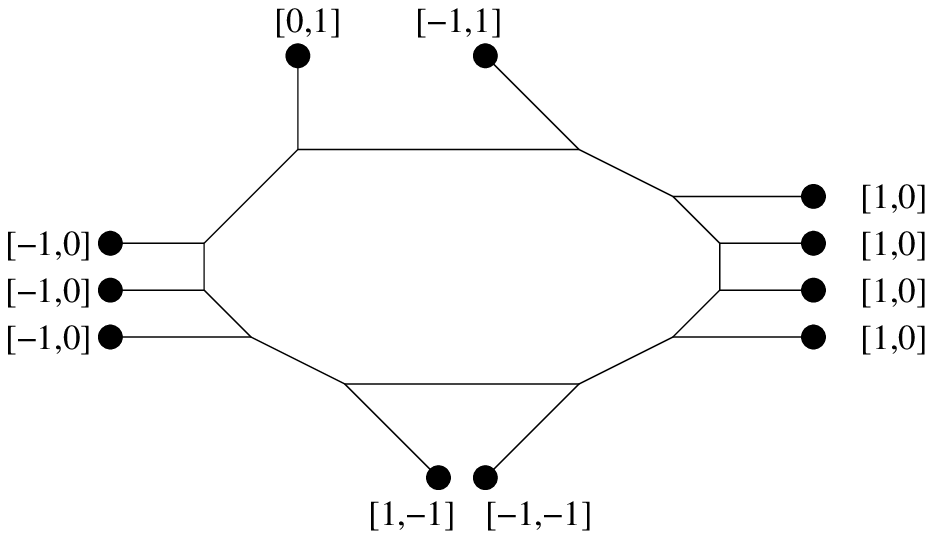}{The web
diagram of local $B_{8}$. } The vanishing 1-cycles are, \bea
C_{1}&=&(-1,-1)\,,\,C_{2}=C_{3}=C_{4}=C_{5}=(1,0)\,,\,C_{6}=(-1,1)\,,\,\\
\nonumber C_{7}&=&(0,1)\,,\, C_{8}=C_{9}=C_{10}=(-1,0)\,,\,
C_{11}=(1,-1)\,. \eea The gauge group is given by, \bea
G=\prod_{a=1}^{11}\mbox{U}(1)_{a}\,. \eea The
corresponding fractional brane charges are, \bea
S_{1}\,:\,\,\mbox{ch}(F_{1})&=&(-1,l-E_{1}-E_{4}-E_{7},1)\\
\nonumber
S_{2}\,:\,\,\mbox{ch}(F_{2})&=&(0,E_{7},-\frac{1}{2})\,,\,\,\\
\nonumber
S_{3}\,:\,\,\mbox{ch}(F_{3})&=&(0,E_{4},-\frac{1}{2})\,,\,\,\\
\nonumber
S_{4}\,:\,\,\mbox{ch}(F_{4})&=&(0,E_{1},-\frac{1}{2})\,,\,\,\\
\nonumber
S_{5}\,:\,\,\mbox{ch}(F_{5})&=&(0,l-E_{3}-E_{6},-\frac{1}{2})\\
\nonumber S_{6}\,:\,\,\mbox{ch}(F_{6})&=&(1,-l+E_{6},0)\\
\nonumber S_{7}\,:\,\,\mbox{ch}(F_{7})&=&(1,-E_{5},-\frac{1}{2})\\
\nonumber
S_{8}\,:\,\,\mbox{ch}(F_{8})&=&(0,-l+E_{3}+E_{5},\frac{1}{2})\\
\nonumber
S_{9}\,:\,\,\mbox{ch}(F_{9})&=&(0,-E_{2},-\frac{1}{2})\,,\,\,\\
\nonumber
S_{10}\,:\,\,\mbox{ch}(F_{10})&=&(0,-E_{8},-\frac{1}{2})\,\,\\
\nonumber S_{11}\,:\,\,\mbox{ch}(F_{11})&=&(-1,E_{2}+E_{8},1)\,.
\eea 
Note that in all the cases we discussed the sum of fractional
branes is equal to a 0-cycle as required by mirror symmetry,
\bea
\sum_{i=1}^{N+3}\mbox{ch}(F_{i})=(0,0,-1)\,.
\eea Also
it is easy to check that the set $\{F_{1},\cdots, F_{N+3}\}$, for
$0\leq N \leq 8$, is an exceptional collection forming a helix on
${\cal B}_{N}$. Another interesting point to note is that since
\bea
\inter{S_{i}}{\sum_{j=1}^{N+3}S_{j}}=0\,,
\eea
therefore it follows that in
corresponding quiver diagram, for each node, the number of
incoming arrows is equal to the number of outgoing arrows. This
statement in terms of gauge theory just states that there are no
chiral anomalies for each of the gauge group factors involved.

\section*{Acknowledgements}
A.I. would like to thank Mina Aganagic, Julie D. Blum, Jacques
Distler, Ansar Fayyazuddin, Tasneem Zehra Husain and Amir-Kian
Kashani-Poor for valuable discussions. A.H. would like to thank
David Berenstein, Emanuel Diaconescu, Bo Feng, Yang-Hui He and
Angel Uranga for valuable discussions. A.H. would also like to
thank the department of Physics at the Weizmann Institute, the
high energy theory group in Tel Aviv University and the ITP in
UCSB for their kind support while completing various stages of
this work. The research of A.I. was supported in part by NSF
grants PHY-0071512. The research of A.H. was supported in part by
the DOE under grant no. DE-FC02-94ER40818, by an A. P. Sloan
Foundation Fellowship, by the Reed Fund Award and by a DOE OJI
Award.

\appendix
\section{Fractional Brane Charges}

In this appendix we explain the calculation of fractional brane
charges. This calculation uses the map the between 3-cycle in the
mirror geometry and vector bundles on del Pezzo surfaces studied
in \cite{HI,MOY2}.

Recall that the geometry we are considering is an elliptic and a
$\bbbc^{\,\times}$ fibration over the z-plane. The
$\bbbc^{\,\times}$ fibration is universal in the sense that for
all cases it degenerates at $z=0$. Thus all the information is
contained in the elliptic fibration over the z-plane and we can
map all the 3-cycles in this geometry to string junctions with
support on 7-branes and a D3-brane. The $(p,q)$ 7-branes
correspond to the degenerate fibers of the elliptic fibration and
the D3-brane corresponds to the degeneration of the
$\bbbc^{\,\times}$ fibration. Thus BPS states of the ${\cal N}=2$
theory obtained from compactification of Type IIB on the mirror
manifold are the same as the BPS states of the $N=2$ theory on the
D3-brane in the background of certain mutually non-local 7-branes.
The 7-brane backgrounds which correspond to the mirror of local
del Pezzo surfaces were studied in detail in \cite{DHIZ} and lead
to broken affine $E_{N}$ gauge symmetry on the 7-branes. \bea
{\cal E}_{N}\,:\,\,\underbrace{[1,0]\cdots
[1,0]}_{N}\,[2,-1]\,[-1,2]\,[-1,-1] \label{egabo}\eea

Here the numbers in square brackets denote the $(p,q)$ charges of
the corresponding 7 branes. The only information we need
concerning the map is that if a string junction $J$ ends on the
D3-brane with charge $(p,q)$ then the rank $r$ and the degree of
the first Chern class $d_{c_{1}}$
\footnote{$d_{\Sigma}=-K_{X}\cdot \Sigma\,,\,\,\,\Sigma\in
H_{2}(X)$, where $K_{X}$ is the anticanonical class of $X$.} are
given by \bea r&=&q\,,\\ \nonumber d_{c_{1}}&=&p-q\,. \label{map1}
\eea Also the self-intersection number of the junction is equal
$-\chi(V,V)$, \bea -J\cdot J=\chi(V,V):=\int_{X}\mbox{ch}(V\otimes
V^{*})\mbox{Td}(X)=r^{2}-c_{1}\cdot c_{1}+2rk, \label{map2} \eea
where $k=\int_{X}\mbox{ch}_{2}(V)$. Eq(\ref{map1}) and
Eq(\ref{map2}) imply that string junctions with support only on a
single [1,0] 7-brane corresponds to the bundle (sheaf) \bea
\mbox{ch}(V_{a})=(0,E_{a},-\frac{1}{2})\,,\,\,\,a=1,\cdots,n
\label{map4}\eea

The bundles corresponding to other string junctions living on $[2,-1]$
, $[-1,2]$ and $[-1,-1]$ can also be obtained using Eq(\ref{map1}) and
Eq(\ref{map2}). Consider the case of string junction on $[n,m]$
7-brane. From Eq(\ref{map1}) it follows that the corresponding bundle
has $ch_{0}=m$. Since $dimH_{2}(B_{0})=1$ with generator $l$ the first
Chern class of the bundle is a multiple of $l$. From $d_{c_{1}}=p-q$
it follows that $c_{1}=\frac{n-m}{3}l$.  The Self-intersection number
of the junction is (since it is a half-sphere) $-1$ and therefore
Eq(\ref{map2}) implies that
$ch_{2}=\frac{n^{2}-2nm-8m^{2}+9}{18m}$. Thus we the following map
\bea [n,m]\,\,&:&\,\mbox{ch}(V)=(m,\frac{n-m}{3}l,
\frac{n^{2}-2nm-8m^{2}+9}{18m})\,,\\ \nonumber
[2,-1]\,\,&:&\,\mbox{ch}(V)=(-1,l,-\frac{1}{2})\,,\\ \nonumber
[-1,2]\,\,&:&\,\mbox{ch}(V)=(2,-l,-\frac{1}{2})\,,\\ \nonumber
[-1,-1]\,\,&:&\,\mbox{ch}(V)=(-1,0,0)\,. \label{map3} \eea Since all
other junctions can be formed by linear combination of these basic
string junctions Eq(\ref{map4}) and Eq(\ref{map3}) provide the
complete map.

\underline{local ${\bf {\cal B}_{1}}$:} As an example we work out the
charges of fractional branes for local ${\cal B}_{1}$.
\onefigure{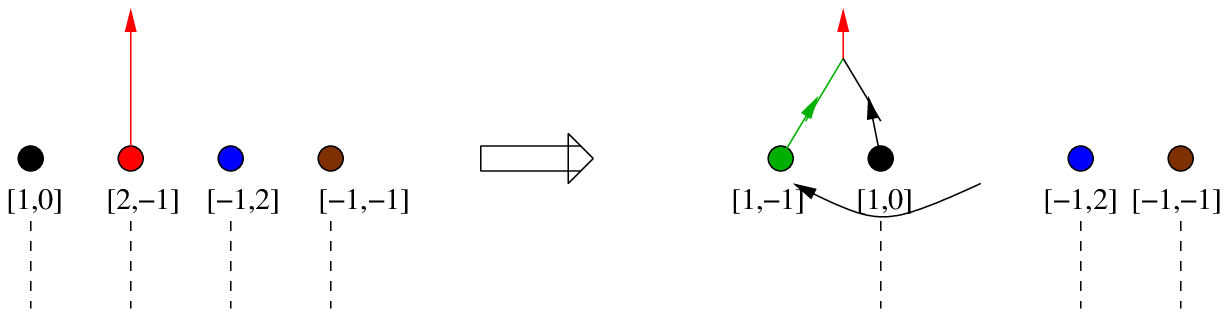}{The branch cut move and the corresponding
action on the string junction.} The branch cut moves which takes
us from the configuration given in equation (\ref{egabo}) to the
one obtained from the toric diagram is shown in \figref{B1-map}.
The sum of junctions on [1,-1] and [1,0] is the same as the
junction on [2,-1], as shown in \figref{B1-map}, therefore \bea
\mbox{ch}(F_{[1,-1]})&=&\mbox{ch}(F_{[2,-1]})-\mbox{ch}(F_{[1,0]})\\
\nonumber &=&(-1,l,-\frac{1}{2})-(0,E_{1},-\frac{1}{2})\\
\nonumber &=&(-1,l-E_{1},0) \eea

Similarly we determine the charge of fractional branes in other local
del Pezzo geometries by starting with the configuration given in
(\ref{egabo}) and converting in into the one given by the toric
diagram using branch cut moves.


\begin{thebibliography}{99}

\bibitem{KKV} S. Katz, A. Klemm, C. Vafa, ``Geometric Engineering of
Quantum Field Theories,'' Nucl. Phys B497 (1997) 173-195,
{\tt hep-th/9609239}.

\bibitem{KMV}
S. Katz, P. Mayr, C, Vafa, ``Mirror Symmetry and Exact Solution of N=2
Theories 4D N=2 Gauge Theories I'', Adv. Theor. Math. Phys. 1
(1998) 53-114, {\tt hep-th/9706110}.

\bibitem{CIV}
F. Cachazo, K. Intriligator, C. Vafa, ``A Large N Duality via a
Geometric Transition,'' Nucl. Phys. B603 (2001) 2-41,
{\tt hep-th/0103067},\\
F. Cachazo, S. Katz, C. Vafa, ``Geometric Transitions and
N=1 Quiver Theories,'' {\tt hep-th/0108120}.

\bibitem{MS}
D.R. Morrison, N. Seiberg, ``Extremal Transitions and
Five-Dimensional Supersymmetric Field Theories,'' Nucl. Phys. B483
(1997) 229-247, {\tt hep-th/9609070}.

\bibitem{DKV}
M. R. Douglas, S. Katz, C. Vafa, ``Small Instantons, del Pezzo
Surfaces and Type I$'$ Theory,'' Nucl. Phys. B497 (1997) 155-172,
{\tt hep-th/9609071}.

\bibitem{IMS}
K. Intriligator, D.R. Morrison, N. Seiberg, ``Five Dimensional
Supersymmetric Gauge Theories and Degeneration of calabi-Yau
Spaces,'' Nucl.Phys. B497 (1997) 56-100, {\tt hep-th/9702198}.

\bibitem{FHH1}
B. Feng, A. Hanany, Y. -H. He, ``D-Brane Gauge Theories from Toric
Singularities and Toric Duality,'' {\tt hep-th/0003085}.

\bibitem{FHH2}
B. Feng, A. Hanany, Y. -H. He, ``Phase Structure of D-brane Gauge
Theories and Toric Duality,'' {\tt hep-th/0104259}.

\bibitem{BGLP}
C. Beasley, B. R. Greene, C. I. Lazaroiu, M.R. Plesser,
``D3-branes on partial resolution of abelian quotient
singularities of Calabi-Yau threefolds,'' Nucl.Phys. B566 (2000)
599-640, {\tt hep-th/9907186}.

\bibitem{HI}
T. Hauer, A. Iqbal, ``Del Pezzo Surfaces and Affine 7-brane
Backgrounds,'' JHEP 0001 (2000) 043, {\tt hep-th/9910054}.

\bibitem{MOY2}
K. Mohri, Y. Onjo, S. K. Yang, ``Duality Between String Junctions
and D-Branes on Del Pezzo Surfaces,'' {\tt hep-th/0007243}.

\bibitem{HI2}
A. Hanany, A. Iqbal, in progress.

\bibitem{papers}
S. Katz, C. Vafa, ``Geometric Engineering of N=1 Quantum Field
Theories,'' Nucl. Phys. B497 (1997) 196-204, {\tt hep-th/9611090},\\
M. Bershadsky, A. Johansen, T. Pantev, V. Sadov, C. Vafa,
``F-Theory, geometric Engineering and N=1 Dualities,'' Nucl. Phys.
B505 (1997) 153-164, {\tt hep-th/9612052},\\ B. Zwiebach, C. Vafa,
``N=1 Dualities of SO and USp Gauge Theories and T-Duality of
String Theory,'' Nucl. Phys. B506 (1997) 143-156,
{\tt hep-th/9701015},\\
J. D. Edelstein, K. Oh, R. Tatar, 
``Orientifold, Geometric Transition and Large N Duality for SO/Sp 
Gauge Theories,'' JHEP 0105 (2001) 009, {\tt hep-th/0104037}.

\bibitem{BL}
D. Berenstein, R. G. Leigh, ``Resolution of Stringy Singularities
by Non-commutative Algebras'',
JHEP 0106 (2001) 030, {\tt hep-th/0105229}.

\bibitem{YY}
Y. Yamada, S. K. Yang, ``Affine 7-brane Backgrounds and
Five-Dimensional $E_{N}$ Theories on $S^{1}$,'' Nucl. Phys. B566
(2000) 642-660, {\tt hep-th/9907134}.

\bibitem{HV}
K. Hori, C. Vafa, ``Mirror symmetry,'' {\tt hep-th/0002002}.

\bibitem{HIV}
K. Hori, A. Iqbal, C. Vafa, ``D-Branes and Mirror Symmetry,''
{\tt hep-th/0005247}.

\bibitem{SZ}
A. Sen, B. Zwiebach, ``Stable Non-BPS States in F-theory,''
{\tt hep-th/9907164}.

\bibitem{DHIZ}
O. DeWolfe, T. Hauer, A. Iqbal, B. Zwiebach, ``Uncovering Infinite
Symmetries on [p,q] 7-branes: Kac-Moody Algebras and Beyond,''
Adv. Theor. Math. Phys. 3 (1999) 1835-1891, {\tt hep-th/9812209}.


\bibitem{webs}
O. Aharony, A. Hanany, B. Kol, ``Webs of $(p,q)$ 5-branes, Five
dimensional Field Theories and Grid Diagrams,'' {\tt hep-th/9710116}.


\bibitem{LV}
N. C. Leung, C. Vafa, ``Branes and Toric Geometry,'' Adv. Theor.
Math. Phys. 2 (1998) 91-118, {\tt hep-th/9711013}.

\bibitem{DHIK}
O. DeWolfe, A. Hanany, A. Iqbal, E. Katz, ``Five-branes,
Seven-branes and Five-dimensional $E_n$ field Theories,''
{\tt hep-th/9902179}.


\bibitem{affine}
O. DeWolfe, ``Affine Lie Algebras, String Junctions and
7-Branes,'' Nucl.Phys. B550 (1999) 622-637, {\tt hep-th/9809026}.

\bibitem{weyl}
T. hauer, A. Iqbal, B. Zwiebach, ``Duality and Weyl Symmetry of
7-brane Configurations,'' {\tt hep-th/0002127}.

\bibitem{vafa}
C. Vafa, ``Superstrings and Topological Strings at Large N,''
{\tt hep-th/0008142}.

\bibitem{DG}
D. -E. Diaconescu, J. Gomis, ``Fractional Branes and Boundary
States in Orbifold Theories,'' JHEP 0010 (2000) 001,
{\tt hep-th/9906242}.

\bibitem{DFR}
M. R. Douglas, B. Fiol, C. Romelsberger, ``The spectrum of BPS
branes on a noncompact Calabi-Yau,'' {\tt hep-th/0003263}.

\bibitem{MOY1}
K. Mohri, Y. Onjo, S.-K. Yang, ``Closed Sub-Monodromy Problems,
Local Mirror Symmetry and Branes on Orbifolds,'' {\tt hep-th/0009072}.

\end{thebibliography}
\end{document}